\documentclass[journal=jpccck,manuscript=article]{achemso}
\usepackage{amssymb}

\newcommand{\g}{graphene} 
\newcommand{\gh}{{\it graphane}} 

\newcommand{\Gh}{{\it Graphane}}

    \author {Bhalchandra S. Pujari}
    \email{bspujari@physics.unipune.ernet.in}
    \author {D. G. Kanhere}
    \affiliation{Department of Physics, University of Pune,  Ganeshkhind, Pune--411
    007, India.}
    \alsoaffiliation{Centre for Modelling and Simulations, University of Pune,
    Ganeshkhind, Pune--411 007, India.}

    \title {Density functional investigations of defect induced mid-gap states in graphane} 
                 
\begin{document}

    \begin{abstract} 
      We have carried out {\it ab initio} electronic structure calculations on {\gh}
      (hydrogenated {\g}) with single and double vacancy defects. Our analysis of the
      density of states  reveal that such vacancies induce the mid gap states and modify
      the band gap.  The induced states are due to the unpaired electrons on carbon atoms.
      Interestingly the placement and the number of such states is found to be sensitive
      to the distance between the vacancies. Furthermore we also found that in most of
      the cases the vacancies induce  a local magnetic moment.
    \end{abstract}

\section{Introduction}

{\Gh}\footnote{In order to avoid the confusion with the word `{\g}', the word `{\gh}' is
written in italics throughout this paper.} {\it i.e.} the hydrogenated {\g},  is a recent
addition to the family of novel carbon materials known for their exotic properties.
\cite{sofo:153401, boukhvalov, casolo} Recently {\gh} is experimentally realized by Elias
{\it et al} \cite{elias01302009} who demonstrated that the process of hydrogenation is
reversible.  This observation makes {\gh} a suitable candidate for the hydrogen-storage
material.  Equally interesting is the possibility of a direct observation of
metal-insulator transition in two dimensional systems \cite{fuhrer}.  as a function of
hydrogen coverage.  {\Gh} is an insulator with a reported theoretical band gap of $\sim$
3.5 eV using density functional theory (DFT) \cite{sofo:153401}.

It is well known that a single sheet of {\g} is susceptible  to a variety of disorders
like topological defects, impurity states, ripples, cracks etc.  Pereira {\it et al}
\cite{pereira2008, pereira} have studied different models of local disorders in {\g} and
have investigated their electronic structure within tight binding method. They have
observed a significant changes in the low energy spectrum of {\g} viz., localized zero
modes, strong resonances, gap and pseudogap behavior etc depending upon the type of
disorder. Their results also indicate that by and large disorder significantly modifies
the states near the Fermi level.  Yazyev and Helm \cite{yazyev} have studied defect
induced magnetism in {\g} using  DFT. Their work shows that
the adsorption of hydrogen or the creation of defect on {\g} sheet lead to the local
magnetic moment.  For an extensive survey of studies of disorders in {\g} we refer the
reader to a recent review by Castro Neto {\it et al} \cite{netormp}.  We wish to point out
that most of the novel properties of {\g} arise due to the nature of the 
density of states (DOS) near the Fermi level and these states are sensitive to the presence of defects.

Quite clearly a detailed study of defect induced states is warranted not only for {\g} but
also for {\gh}.  So far there are no reports of systematic investigations of the effects
of such defects on the properties of {\gh}.  In the present work we  focus on the
electronic structure of {\gh} with topological defects created by the removal of one and
two carbon atoms. Such defects are experimentally realized using high-energy ion
beams as demonstrated by  Jin {\it et al} \cite{jin:205501} by creating a stable carbon
chain from {\g} sheet.

The present work is based on  spin density functional theory (SDFT), which is known to
underestimate the band gap.  A recent calculation by Leb\`egue {\it et al} \cite{lebegue}
based on GW approximation estimated the band gap of {\gh} to be 5.4 eV.  The authors have
also shown that the removal of a single hydrogen atom produces mid gap states.

The paper is organized as follows. We present relevant computational details in section
\ref{sec:compdet}. In section \ref{sec:prisine}, the results for pristine {\gh} is
summarized for the purpose of comparison. The main results of the electronic structure
calculations on the single and the double vacancy defects are presented in sections
\ref{sec:single} and \ref{sec:double} respectively. Finally the conclusions are presented
in section \ref{sec:concl}.

\section{Computational Details \label{sec:compdet}}

All the calculations have been performed on a monolayer {\gh}, having geometry as
described by Sofo {\it et al} \cite{sofo:153401}. We have used plane wave based DFT as
implemented in Quantum Espresso\footnote{http://www.quantum-espresso.org/} and VASP
\cite{vasp} using the generalized gradient approximation (GGA) \cite{PBE,PBEerr} for
exchange-correlation potential.  For single vacancy study the primitive {\gh} cell is
repeated 5 times in $X$ and $Y$ directions while for that of double vacancies the
repetitions are of 7 units. When creating the vacancy, the carbon and the attached hydrogen
are removed. After removing two (four) atoms to form single (double) defect(s) there are
total of 98  (192) atoms on the {\gh} plane.  The vertical axis ($Z$) of the cell is kept
as large as 10 {\AA}  to avoid the interactions between the {\gh} sheets.  The energy and
force thresholds are kept at 10$^{-6}$ eV and 10$^{-5}$ eV/{\AA} respectively.  It may be
mentioned that even though the unit cell is large it was found to be necessary to use  5
$\times$ 5 Monkhorst-Pack K-grid for acceptable convergence in energy during the
optimization and the self consistency. After optimization we have used 11 $\times$ 11
Monkhorst-Pack K-grid for final calculations of DOS and other quantities.

\section{Results and discussion}

\subsection{Pristine {\Gh} \label{sec:prisine}} 

Before we present the results of the vacancy studies, it is instructive to summarize the
properties of a pure {\gh}.  Our results on pristine {\gh} are consistent with earlier
reports \cite{sofo:153401, boukhvalov}. {\Gh} is known to have two distinct conformations
depending upon the position of hydrogen atoms with respect to {\g} plane. In the {\it
chair} conformer the hydrogen atoms are attached to carbon atoms in alternating manner to
both the sides of the plane while in the {\it boat} conformer the pairs of hydrogen atoms
are attached in alternating manner \cite{sofo:153401}. Out of these two the chair
conformer is energetically more favorable, hence in the present work we have studied the
chair conformer only.  It is interesting to note that in {\g} the $K$ point of Brillouin
zone is degenerate (no gap in DOS) however for {\gh} the minimum gap is observed at
$\Gamma$ point ($\sim$ 3.5 eV) 
while $K$ point develops rather large gap $\sim$ 12 eV.

\begin{figure}
  \begin{center}
    \includegraphics[width=8cm]{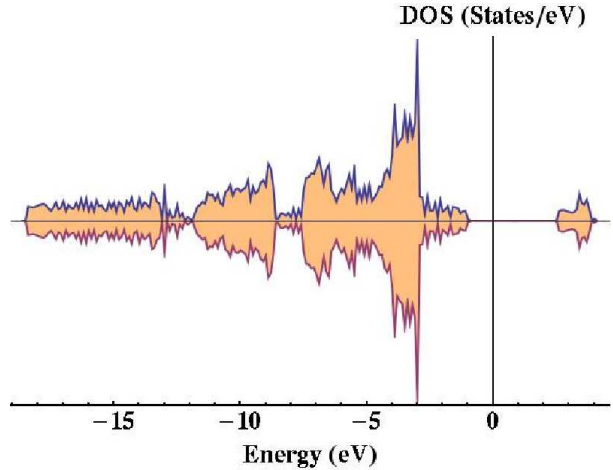}
  \end{center}
  \caption{Spin polarized total DOS for pristine {\gh}. 
  The upper and lower halves of the graph represent up and down spins respectively.  The
  Fermi energy is taken at 0 and marked by a solid vertical line.  The system is
  non-magnetic with a band gap of $\sim 3.5 $ eV.  \label{fig:puregraphanedos}
  }
\end{figure}

We now discuss the DOS of pristine {\gh} which is shown in Figure
\ref{fig:puregraphanedos}. The Fermi energy is taken at 0 and is marked by solid vertical
line.  The largest peak seen (at $\sim$ -3 eV) is due to the peculiar $sp^3$-like bonding
between carbon and hydrogen.  The states  at the top of the valance band ($\sim$ -1 eV)
are mainly comprises of $p$ states forming to the $\sigma$ bonds among the carbon atoms.
Unlike {\g} there are no $\pi$ bonds in {\gh}.

\subsection{Single vacancy  \label{sec:single}}

\begin{figure}
  \center{
  \includegraphics[width=7cm]{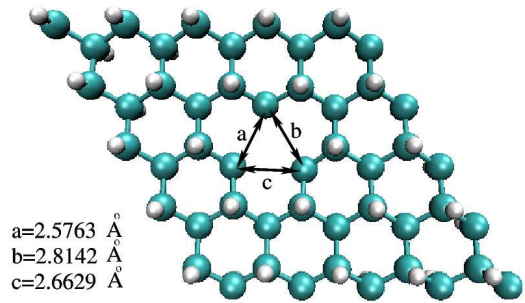}
  }
  \caption{Optimized structure of {\gh} with a single vacancy. Carbon atoms are shown in
  cyan color while hydrogen atoms are in white.  The distorted nature of the sublattice
  surrounding the vacancy is clearly evident (marked by the triangle).  Pristine
  sublattice is an equilateral triangle with $a=2.52$ \AA. \label{fig:GraphaneSingle}}
\end{figure}

We begin our discussion by presenting the study of a single vacancy and its effects on the
geometry, electronic structure  of {\gh}.  A single vacancy is created by removing one
carbon atom along with the attached hydrogen from pure {\gh}.  The fully relaxed structure
is shown in figure \ref{fig:GraphaneSingle}. The immediate effect of the removal of atoms
is seen on the sublattice surrounding the vacancy (indicated by the triangle in the
figure). In an ideal {\gh} the carbon sublattice is an equilateral triangle with length of
the side $a=2.52$ \AA, however in the presence of vacancy the sublattice deforms
substantially. Due to reduced coordination number, three carbon atoms on the triangle are
pushed away from each other and remain as a part of the hexagon. The distortion of the
triangle is asymmetric and the symmetry breaking is due to Jahn-Teller effect.  The
deformation seen here is qualitatively different than seen in the case of {\g} where  two
of the carbon atoms move close to each other to form a $\sigma$ bond \cite{yazyev}. We do
not see formation of bonds between the carbon atoms.  Thus vacancy leads three dangling
bonds on the triangle.  As a consequence, the system becomes magnetic with magnetic moment
1 $\mu_B$.  As we shall see these unpaired electrons  have an interesting consequences
on the DOS. 

\begin{figure}
  \center{
  \includegraphics[width=8cm]{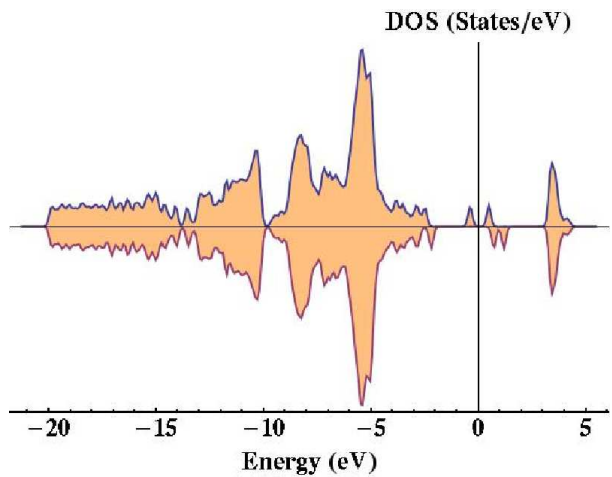}
  }
  \caption{DOS of {\gh} with single vacancy displaying the induced states in the gap. 
  Fermi energy is at zero and shown by the vertical line. The induced states are partially
  occupied for up electrons. \label{fig:singledos}}
\end{figure}

Figure \ref{fig:singledos} shows the spin polarized DOS displaying some remarkable
features. The DOS with single vacancy is substantially different from that of pristine
{\gh} \cite{sofo:153401} especially near the Fermi energy (marked by a vertical line in
the figure). Clearly the vacancy has induced the {\it mid gap} states which are partially
occupied.  In particular  the induced spin up states are partially occupied while the
induced spin down states are completely empty.  The appearance of mid gap states can be
attributed to the unpaired electrons from three carbon atoms due to which the states are
pushed up from valance band. To ascertain this observation we have hydrogenated three
available dangling bonds. Our results show that the increasing hydrogen concentration
(from one to three) steadily reduces the density of induced states (figure not shown).
Finally with complete hydrogenation, induced states vanish and the band gap reduces to
$\sim 3$ eV.

\begin{figure}
  \center{
  \includegraphics[width=8cm]{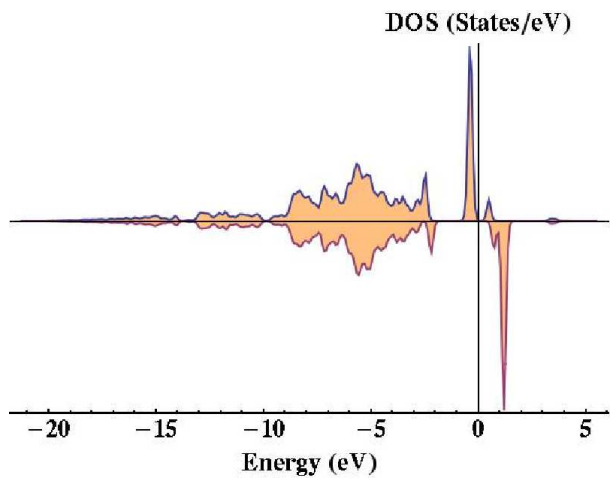}
  }
  \caption{The site projected density of states (PDOS) on one of the carbon atom
  surrounding the vacancy. It is evident that most of the contribution to the induced
  states come from the carbon atoms on the sublattice.
  \label{fig:pdos}}
\end{figure}

That the induced states are due to the unpaired electrons can be further seen from
examination of the site projected DOS. Figure \ref{fig:pdos} shows the site projected DOS
(PDOS) on one of the carbon atoms of the triangle.  It can be seen that almost all the
contribution to the mid gap states come from the three carbon atoms. In fact the
contribution from carbon atoms except the three on the sublattice is negligible.
Interestingly the hydrogen atoms associated with carbon atom on the triangle do not give
any significant contribution.  Furthermore the states turned out to be dominantly
$p$-like.

\begin{figure}
  \center{
  \includegraphics[width=6cm]{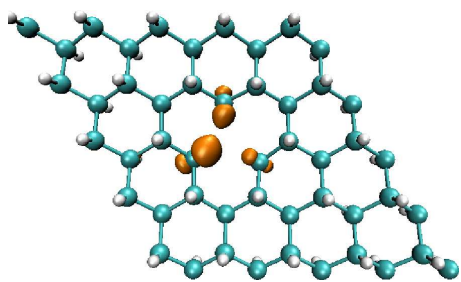}
  }
  \caption{Isosurface of the charge density of the occupied state just below the Fermi
  level for the case of single vacancy. The state is dominantly $p$-like and is highly localized.
  \label{fig:fermiorbital}}
\end{figure}

Although there are states near the Fermi energy, these states need not necessarily
conduct.  Figure \ref{fig:fermiorbital} shows the isosurface (at one fifth of the maximum
value) of the state just below Fermi level.  This state is induced due to the vacancy and
can be seen to be localized.  A careful examination of the isosurfaces at the different
values reveals that the induced state extends up to three to four nearest neighbouring
carbon site. Thus the conduction mechanism through such orbitals may possibly be by
hopping.

\subsection{Double Defect \label{sec:double}}

The results of a single vacancy gives enough impetus to study the effects of multiple
vacancies, especially on the mid gap states. Indeed, as we shall demonstrate, the nature
and the placement of induced states are sensitive to the vacancy-vacancy interaction.
This aspect brings in an interesting possibility of controlling the effective band gap via
creation of vacancy defects. 

For the purpose of this study we have used a larger supercell consisting of 192 atoms.
Several possible scenarios emerge depending upon the separation of the two vacancies.  We
have examined four structures obtained by removing 1) the closest carbon pair (separation,
$D$ = 1.51 \AA), 2) the nearest carbon atoms of the same sublattice ($D$ = 2.52 \AA), 3) a
pair from different sublattice ($D$ = 3.87 \AA) and 4) two carbon atoms having large
separation ($D$ = 10.71 \AA).  In each of the cases we have fully relaxed the structure
and have examined the energetics. We have found that the total energy  is the lowest when
the vacancies are at the closest distance (case 1). The total energies with respect to the
energy of the lowest energy system are shown in Table \ref{tab:deltae}.

\begin{table}
  \centering
  \begin{tabular}{c c c c c}
    \hline
     Separation $D$ & 1.51 \AA & 2.52 \AA & 3.87 \AA & 10.71 \AA\\ 
     \hline
     $\Delta_E$ (eV/atom)& 0.0 & 0.0906 & 0.0185  & 0.0338  \\
     \hline
  \end{tabular}
  \caption{Energy difference ($\Delta_E$) with respect to lowest energy system ({\it i.
  e.} vacancies in vicinity, $D=1.52$ \AA).  $D$ indicate the separation between the
  vacancies.
  \label{tab:deltae}}
\end{table}

\begin{figure}
  \begin{center}
    \includegraphics[width=5cm]{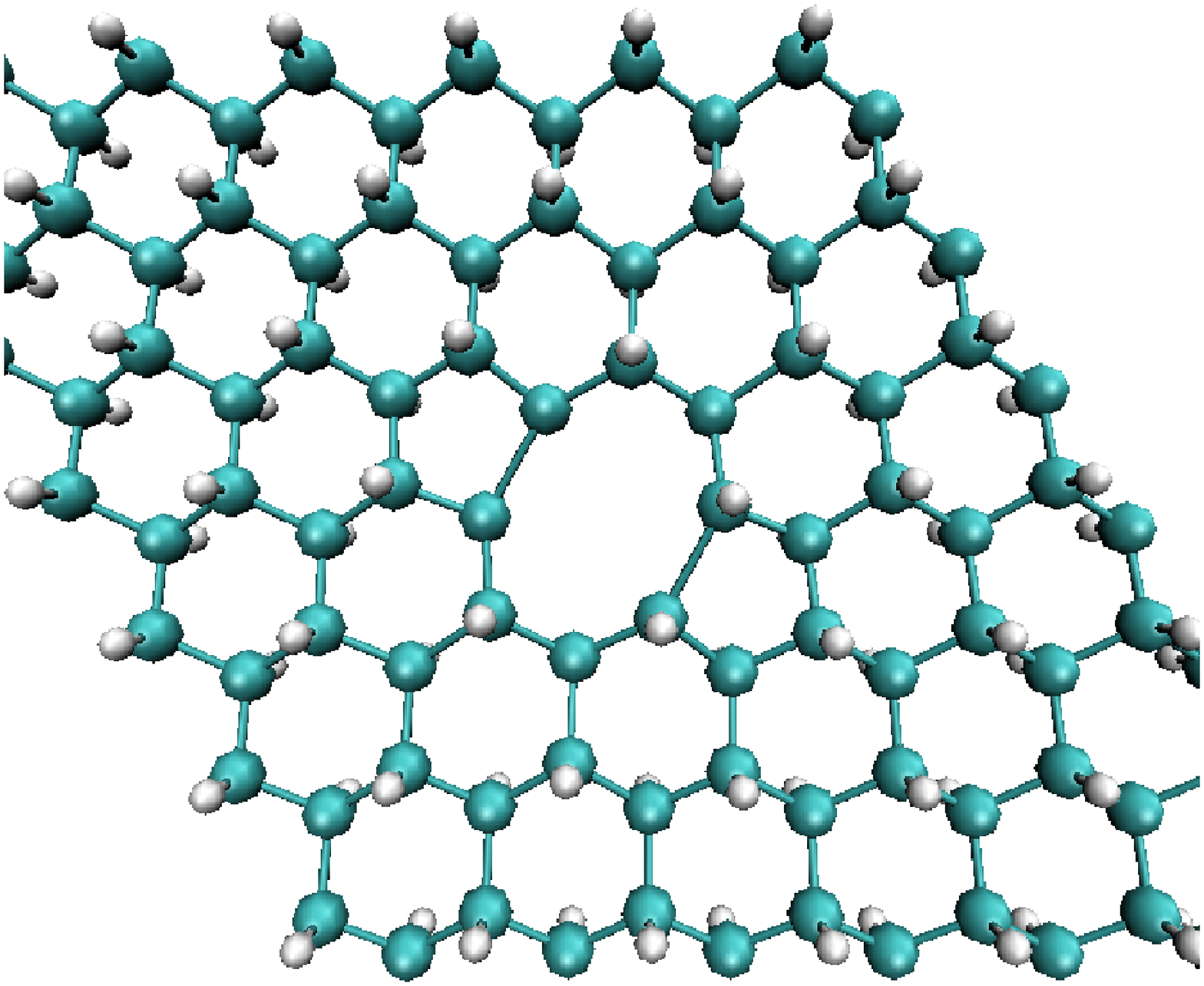} \\
    (a)\\
    \includegraphics[width=7cm]{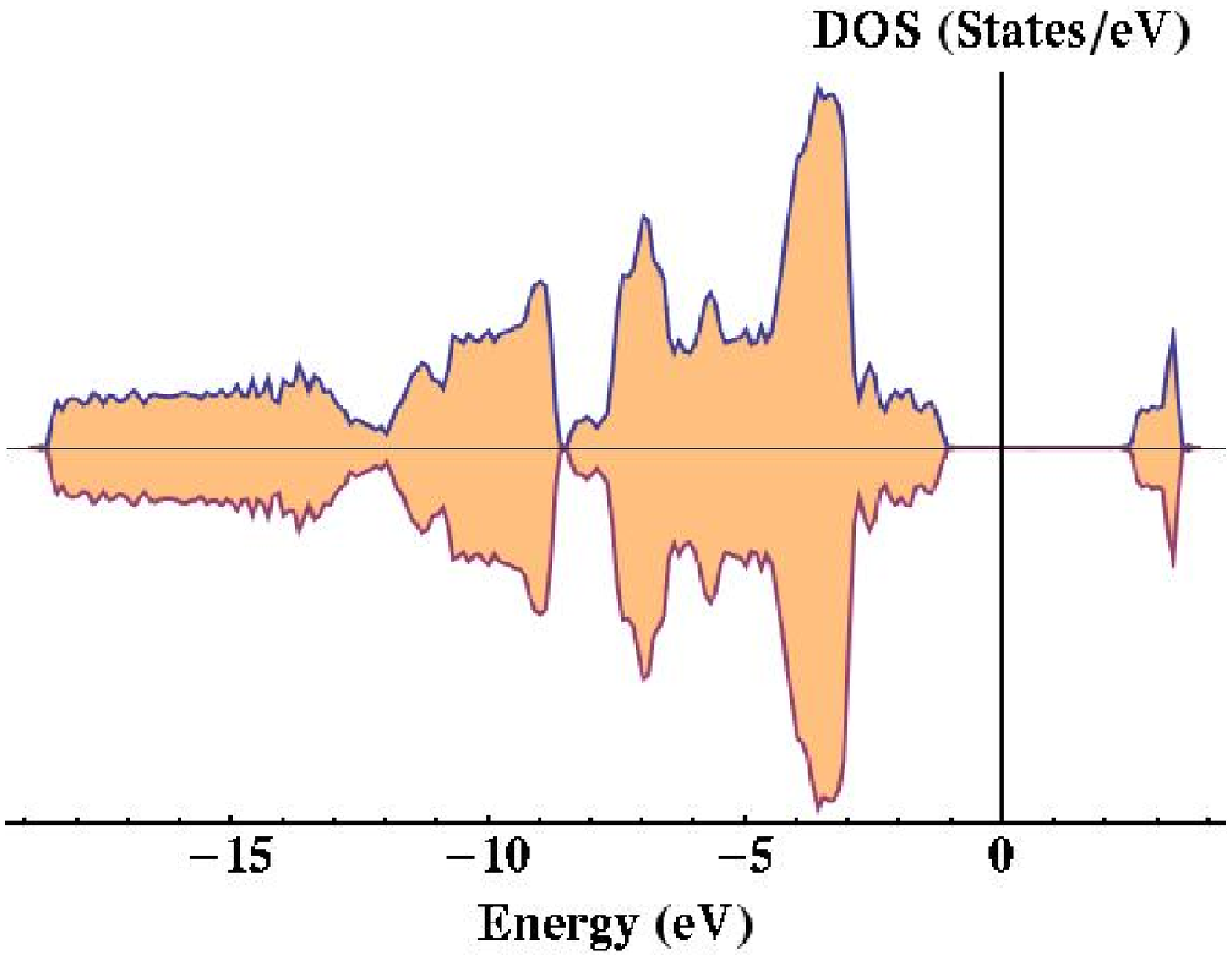}\\ 
    (b)
  \end{center}
  \caption{(a) Optimized structure of {\gh} with two vacancies for the case 1 (see text). 
   The rearrangement of the carbon atoms around the  vacancies can be clearly seen.
  (b) Spin polarized DOS corresponding to structure in (a). The DOS do not show any
  induced mid gap states. Due to the rearrangement of the atoms the band gap is reduced by 0.5
  eV. }
  \label{fig:closedouble}
\end{figure}

Figure \ref{fig:closedouble} shows the optimized structure and DOS for the case 1.
Remarkably, the carbon atoms surrounding the vacancies rearrange to form two new $\sigma$
bonds leading to the formation of 5-8-5 ringed structure.  This peculiar structure does
not leave any unpaired electrons (dangling bonds) unlike the case for single vacancy.  This
is consistent with the observation that there are  no mid gap states in the DOS as can be
seen from figure \ref{fig:closedouble} (b). In this case the band gap is reduced by 0.5 eV
with respect to pristine {\gh}.

\begin{figure}
  \begin{center}
    \includegraphics[width=4cm]{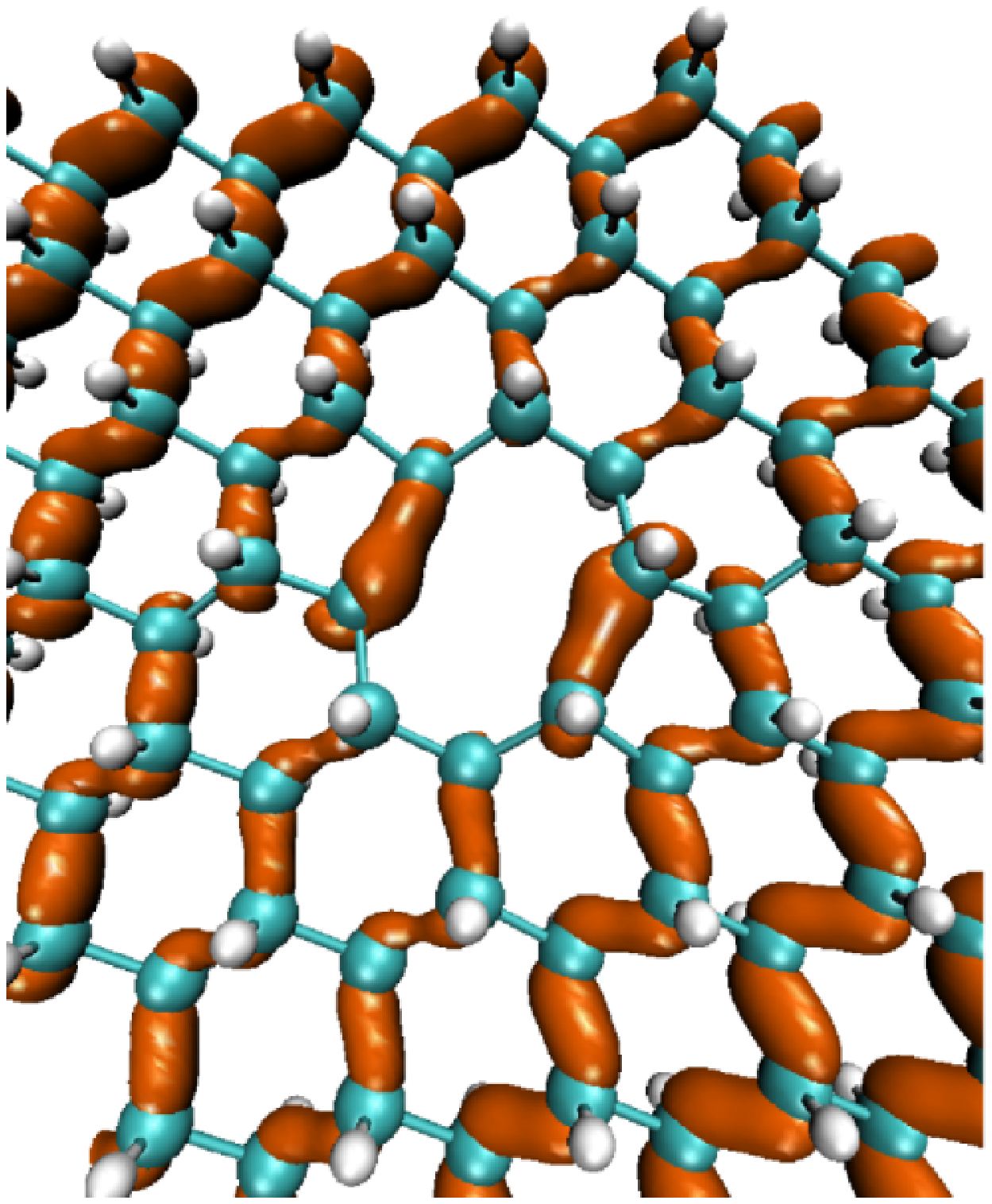}\\
    (a)\\~\\
    \includegraphics[width=5cm]{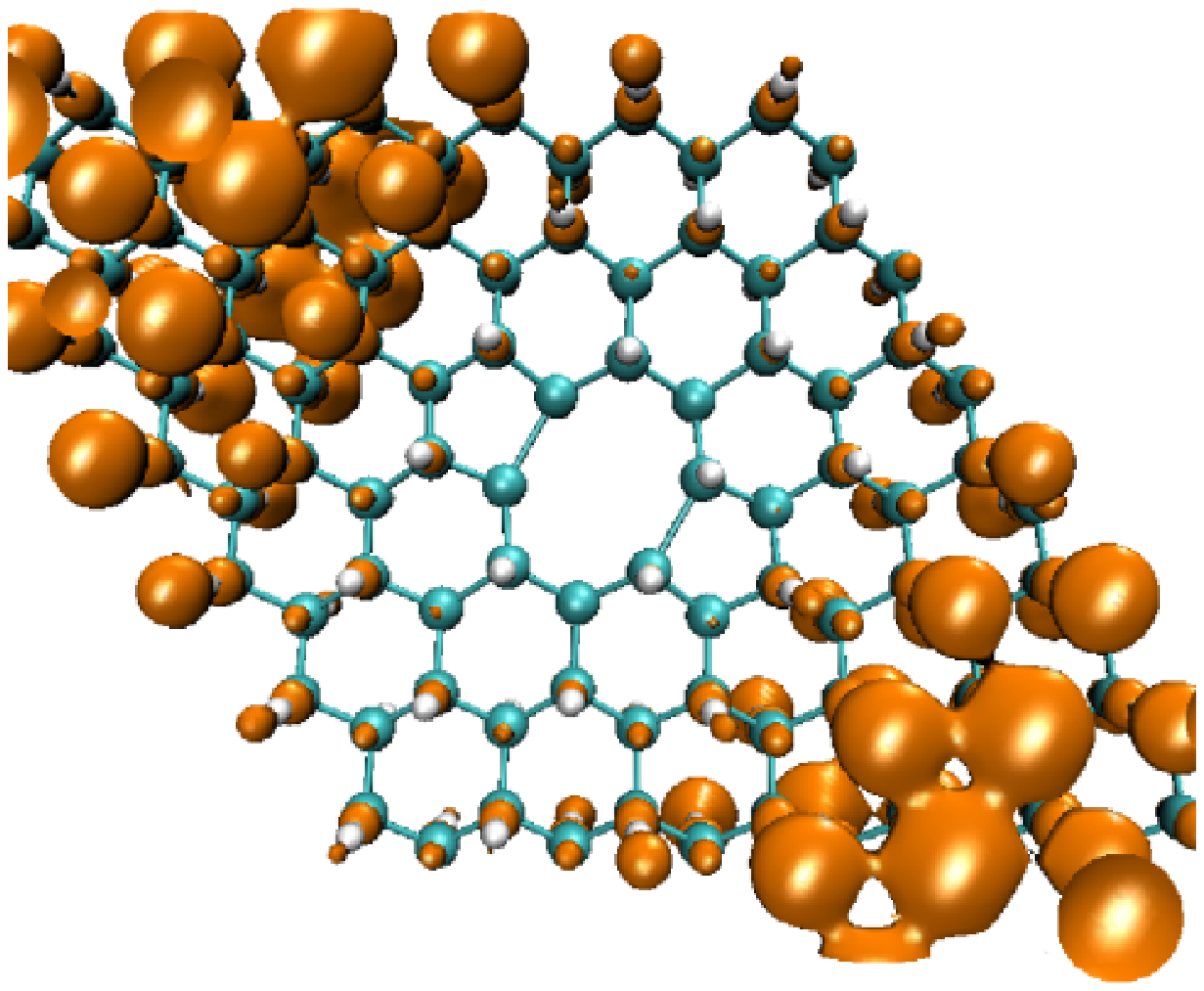}\\
     (b)
  \end{center}
  \caption{
  (a) The charge density of the state at the top of valance band.  Two $\sigma$ bonds
  formed by rearrangement of carbon atoms are clearly seen.
 (b) Highly delocalized state at the bottom of the conduction band. Both the
  states resemble those of pristine {\gh} except for the modifications around the defect
  site.  \label{fig:doubledoscloseopti}}
\end{figure}

In order to ascertain the formation of bonds we have  examined the charge densities of
relevant states. Figure \ref{fig:doubledoscloseopti} (a) and \ref{fig:doubledoscloseopti}
(b) show the charged densities as isosurfaces for two states: one at the top of the
valance band and other at the bottom of the conduction band. The isosurfaces are shown at
one tenth of their maxima.  The charge density of the state in figure
\ref{fig:doubledoscloseopti} (a) which corresponds to the valance band (occupied), clearly
depicts the formation of $\sigma$ bonds among the carbon atoms. It is interesting to note
that the bands at the bottom of the conduction band (figure
\ref{fig:doubledoscloseopti}(b)) are highly delocalized with almost no charge around the
vacancy site.

\begin{figure*} 
  \centering{\includegraphics[width=7cm]{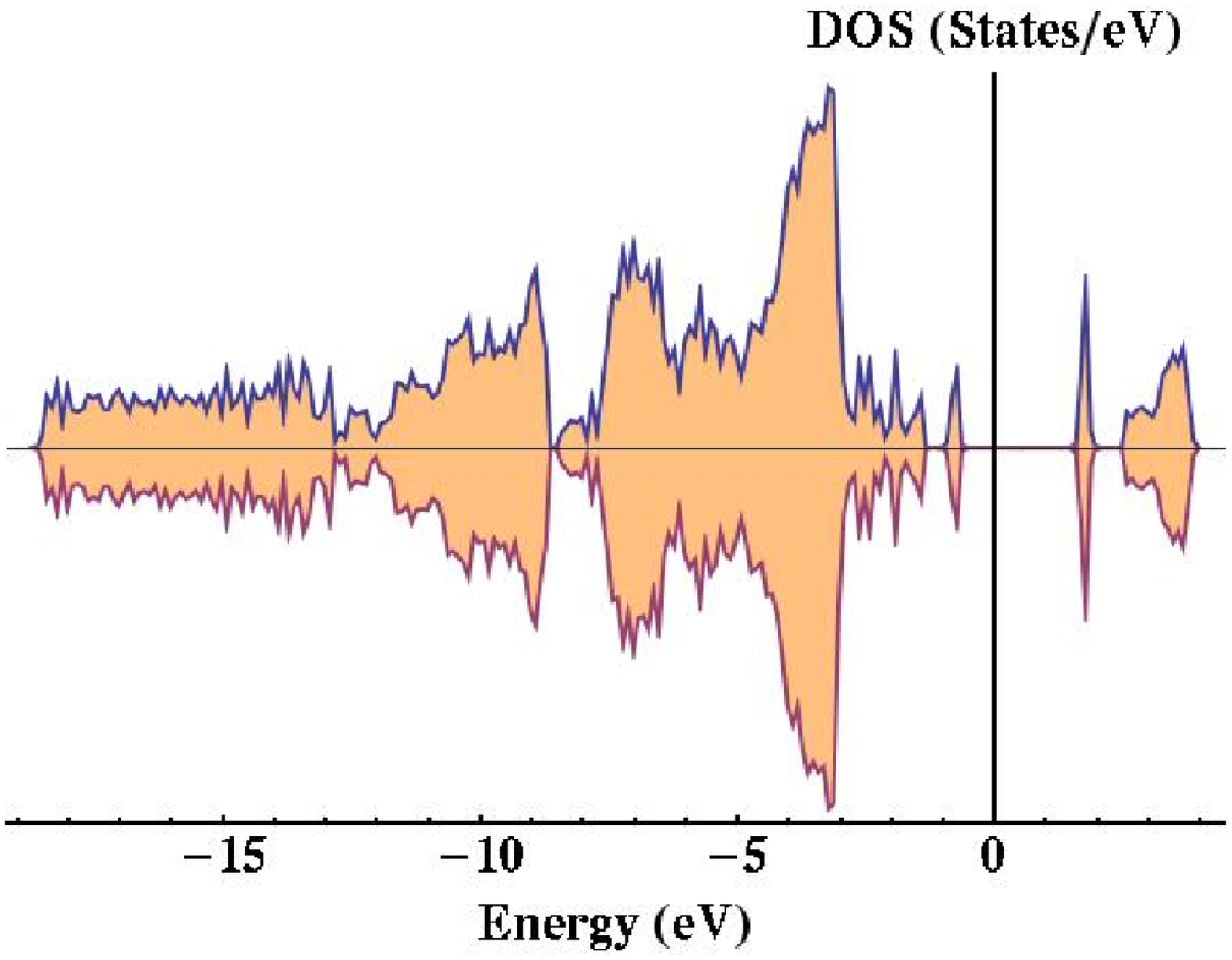}\\
  \centerline{(a): 2.52 \AA.} ~\\~\\
  \includegraphics[width=7cm]{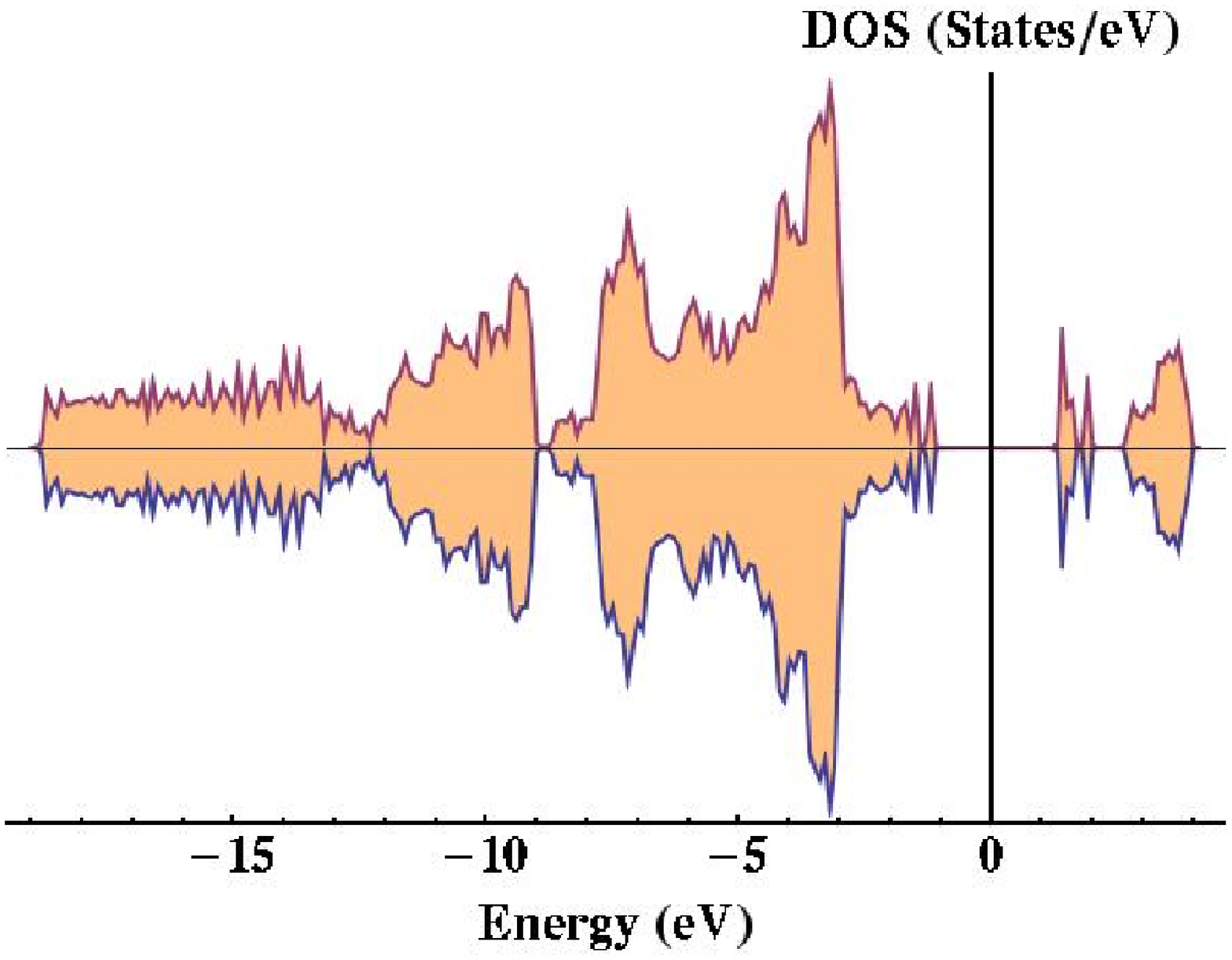} \\
  \centerline{(b): 3.87 \AA.} ~\\~\\
  \includegraphics[width=7cm]{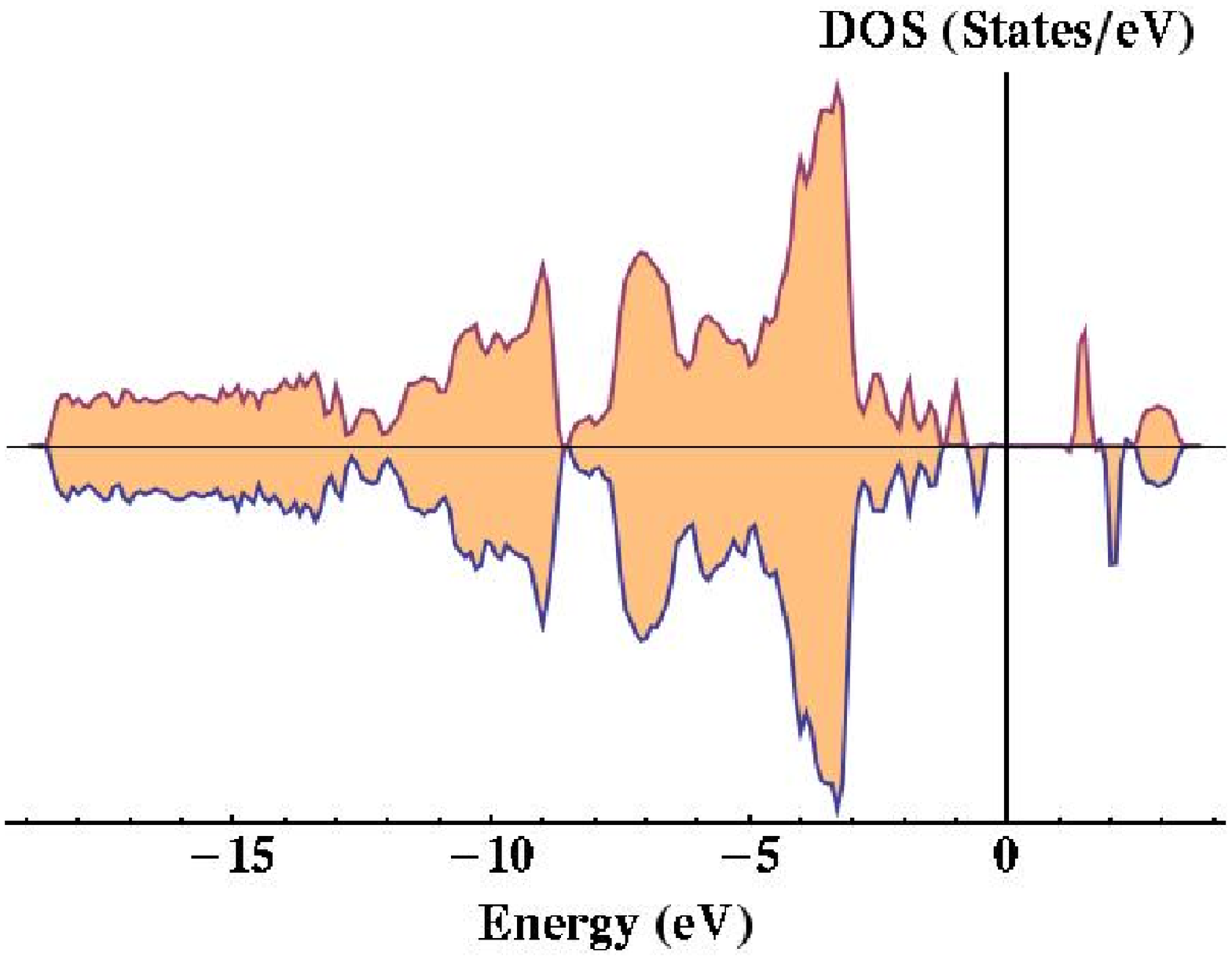}\\
  \centerline{(c): 10.71 \AA}}
  \caption{The DOS of {\gh} with two vacancies for the cases (2), (3) and (4) (see text).
  For each of the cases the nature of the induced states are different.
  (a) The induced states have appeared  near the  conduction  band for the vacancy
  separation of 2.52 \AA.
  (b) For the vacancy separation of 3.87 {\AA} the induced states are broadened,
  effectively reducing the gap from valance band.
  (c) A peculiar magnetic system with 2 $\mu_B$ is seen for the vacancy separation as large as 10.71
  \AA. The induced states are near conduction band and the states on the top  of valance band
  are substantially deformed.  \label{fig:doubledos}} 
\end{figure*}

A different scenario emerges when the vacancies are separated from each other. Now there
are induced states in the gap.  Figure \ref{fig:doubledos} shows the DOS for the three
cases with the vacancy separation of (a) 2.52 {\AA}, (b) 3.87 {\AA} and (c) 10.71 {\AA}
respectively.  Quite clearly in the case of well  separated vacancies, the atoms are not
able to  rearrange so as to form any bonds. This leaves six unpaired electrons.  These
unpaired electrons give rise to the mid gap states typically just above the valance band
and just below the conduction band.  Interestingly their placement is sensitive to  the
separation between the vacancies.  For example, figure \ref{fig:doubledos} (a) shows the
DOS for the vacancy separation $D=$ 2.52 {\AA}. The induced states are seen at about 0.5
eV below the conduction band along with the additional modifications at the edge of the
valance band.  As the separation increases to 3.87 {\AA} (figure \ref{fig:doubledos} (b))
the induced states are broadened with a width of about 1 eV and split in to two distinct
peaks.  In most of the cases the unpaired electrons tend to cancel the spin giving rise to
a non-magnetic states.  However for a large separation {\it i. e.} in the limit of weakly
interacting vacancies, both the vacancies carry a magnetic moment of 1 $\mu_B$. Figure
\ref{fig:doubledos} (c) shows the DOS for system with vacancy separation 10.71 {\AA},
depicting the spin polarized features.

\begin{figure*}
  \begin{center}
    \includegraphics[width=7cm]{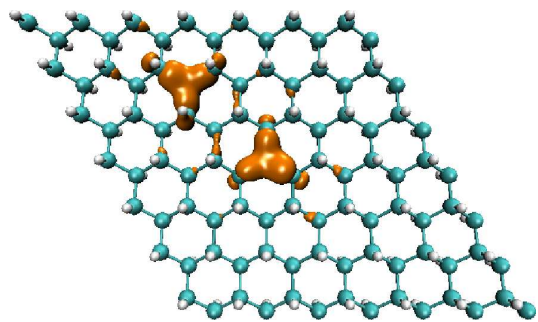}
  \centerline{(a)} ~\\~\\
    \includegraphics[width=7cm]{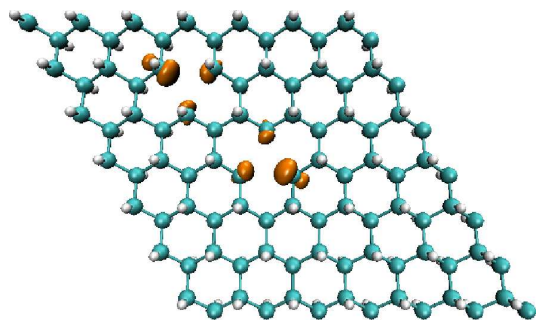}\\
    (b)
  \end{center}
  \caption{The charge densities corresponding to innduced states of {\gh} with two
  vacancies
  separated from each other (case 2). (a) The state at the top of the valance band.  (b)
  The state below the conduction band.  The isosurface is shown at one tenth of its
  maximum. 
  \label{fig:doubleorbital}}
\end{figure*}

The induced states are  seen to be localized both in the case of single as well as double
vacancies. Figure \ref{fig:doubleorbital} depicts the charge density isosurfaces
corresponding two typical induced states for the case 2 ($D=3.87$ \AA).  Figure
\ref{fig:doubleorbital} (a) corresponds to the state at the top of the valance band which
is occupied while figure \ref{fig:doubleorbital} (b) corresponds to the unoccupied state
below the conduction band.  Both the charge densities are plotted at a lower  value of
isosurfaces (one tenth of their maxima). Clearly the states are dominantly $p$-like and
are localized. At this value of isosurface occupied states (figure \ref{fig:doubleorbital}
(a)) is seen to have a weak overlap among the carbon atoms, however at the higher values
the lobes are disconnected.  On the other hand the charge density of the unoccupied state
(figure \ref{fig:doubleorbital} (b)) is highly localized even at one tenth of the maxima.

Apart from the four cases discussed above, we have also calculated the electronic
structure and  DOS for a few more vacancy separations. The results indicate that the
sublattice plays no role in appearance on mid gap states.

\section{Conclusions \label{sec:concl}}

We have carried out an {\it ab initio} investigation of the electronic structure of {\gh}
with single and double vacancy defects. For both the cases, we have analysed the fully
optimized structure, and have examined the  DOS and charge densities as a function of
separation of the vacancies. Our calculations show that the most stable structure is
obtained when the vacancies are adjacent to each other, is accompanied  by the reduction
of band gap. In this case there are no induced mid gap states. However, separated
vacancies induce mid gap states and interestingly their position and width are sensitive
to the vacancy separation. Examination of the charge densities of the induced states show
that these states are localized.  Our calculation brings out the possibility of
manipulating the band gap and the nature of the  mid gap states with the aid of 
vacancy defects in {\gh}.

\begin{acknowledgement}
  B.S.P. would like to acknowledge  CSIR, Govt. of India for financial support (No:
  9/137(0458)/2008-EMR-I). It is a pleasure to acknowledge Center for Development of
  Advanced Computing for computational resources.  Some of the figures are generated by
  using VMD software \cite{vmd}. 
\end{acknowledgement}



\end{document}